# Ab initio analysis for initial process of Joule heating in semiconductors

Emi Minamitani[1,2,3]

1. Institute for Molecular Science, Okazaki 444-8585, Japan

2. JST, PRESTO, Okazaki 444-8585, Japan

3. The Graduate University for Advanced Studies, Okazaki 444-8585, Japan

Abstract

To investigate the initial process of Joule heating in semiconductors microscopically and quantitatively, we developed a theoretical framework for the ab initio evaluation of the carrier energy relaxation in semiconductors under a high electric field using a combination of the two-temperature model and the Boltzmann equation. We employed the method for bulk silicon as a typical example. Consequently, we found a remarkable difference in the energy relaxation processes of the electron and hole carriers. The longitudinal acoustic and optical phonons at the zone boundary contribute to the energy relaxation of electron carriers, whereas they contribute negligibly to that of the hole carriers. In addition, at the band edge, the energy relaxation rate is maximized for the electron carriers, whereas it is suppressed for the hole carriers. These differences stem from the presence/absence of intervalley scattering processes and isotropic/anisotropic band structures in electrons and holes. Our results lay the foundation for controlling the thermal generation in semiconductors by material design.

1. Introduction

Joule heating in semiconductors is a fundamental problem in solid-state physics. The resulting thermal damage, reduction in operational reliability, and power consumption of nanoscale and thin-film transistors have become increasingly critical as the device size decreases. Macroscopically, Joule heating has been simulated as the dot product of the electric field and the current density. The microscopic process of Joule heating in a steady state is described as the energy balance between the electron and phonon intermediated by electron-electron, electron-impurity, electron-phonon, and phonon-phonon interactions under nonequilibrium conditions in the presence of a high electric field.

These scattering processes and the resulting energy relaxation of carriers under a high electric field have long been a central problem in semiconductor physics. Although few studies have been directly concerned with the Joule heating, a number of theoretical and experimental studies have been carried out to explain the high-field effect represented by the saturation of the drift velocity [1–3]. With regard to theoretical methods, one direction is the numerical evaluation of the nonequilibrium distribution function described by the Boltzmann equation. This is often carried out by Monte Carlo simulation which incorporates the acceleration of carriers by an electric field and representative scattering processes. Recently, by conducting a full-band Monte Carlo simulation, Pop *et al.* [4] found that electron energy is mostly transferred to phonons with a low group velocity. Fischetti *et al.* [5] revealed that acoustic phonons significantly contribute to the energy relaxation process. This contradicts the commonly accepted mechanism that most of the energy relaxation is caused by optical phonons. These previous studies provide insight into the complex nature of the energy relaxation from electrons to phonons; however, the empirical treatment of the band structure and the electron–phonon coupling prevents us from understanding the relaxation process quantitatively and microscopically.

The other direction we propose in this paper enables the quantitative investigation of the energy relaxation process with the ab initio treatment of electronic states and electron-phonon interactions by approximating the nonequilibrium electron distribution function as being in equilibrium with an effective temperature. This approximation is known as the two-temperature model because the electrons and phonons are described by different effective temperatures, the electron temperature ($T_e$) and lattice temperature ($T_l$). This model has been widely applied to analyze the carrier relaxation process in semiconductors and other materials [6–16]. The two-temperature model becomes reasonable if the electron equilibration time is sufficiently shorter than the time required for energy relaxation from electrons to phonons. Under such conditions, the Joule heating process can be described by the following three steps: First, electrons accelerated by an electric field are scattered elastically by electron-electron and electron-impurity interactions with a short relaxation time. This process randomizes the electron energy and momentum, and consequently, the electron distribution becomes isotropic in **k**-space, which can be described by a Fermi distribution with $T_e$. Second, the inelastic electron-phonon scattering with a longer energy relaxation time occurs, and the energy is transferred from the hot electrons to the cold phonons specified by $T_l$. Finally, the excited phonons are thermalized by the slow phonon–phonon interaction, and the heat energy is radiated to the environment or transported to a thermal bath such as a substrate and/or electrode by thermal phonons. We focus on the second energy transfer process as the initial step of Joule heating and develop a method to evaluate this process on the basis of ab initio calculations.

Recent progress in ab initio calculations by combining density functional theory (DFT) and the Wannier interpolation technique [17–20] has enabled the evaluation of electron–phonon coupling with high precision. This technique has been employed to evaluate the transport properties of semiconducting materials [21–30]. Here, we combine the ab initio calculation of the transport properties with the two-temperature concept. By considering the phonon Boltzmann equation under the two-temperature model, we formulate the carrier energy relaxation rate as a function of $T_e$, as explained in Sec. IIA. In addition, we evaluate the mobility as a function of $T_e$ with fixed $T_l$. Using the balance between the energy relaxation rate and macroscopic Joule heating obtained from the two-temperature mobility, the applied electric field ($E$) equivalent to $T_e$ is determined as explained in Sec. IIB. The details of the DFT calculations are summarized in Sec. IIC. We applied this theoretical approach to bulk Si as a specific target. The results of the conversion between $T_e$ and $E$ in bulk Si are presented in Sec. IIIA. Then, we investigated the dependence of the carrier energy relaxation rate on $E$ and the microscopic details of the phonon modes that contribute to the relaxation process. The results are summarized in Sec. IIIB. We found a remarkable difference in the energy relaxation processes of the electron and hole carriers. The longitudinal acoustic and optical phonons at the zone boundary contribute to the energy relaxation process of the electron carriers, whereas they negligibly contribute to that of the hole carriers. Moreover, the distributions of the relaxation rate with respect to the carrier energy are different for the electron and hole carriers. The energy relaxation rate of an electron carrier is maximized at the conduction band minimum (CBM), whereas for a hole carrier, this occurs at energy slightly shifted from the valence band maximum (VBM). These variations can be attributed to the different electron–phonon scattering processes and band structures of the electron and hole carriers. These analyses provide us with the quantitative carrier energy relaxation rate as a function of $E$, and the amount of the contribution from the respective electron-phonon interaction process to the initial process of Joule heating.

II. Methods

A. Derivation of energy relaxation rate

In the steady state, the excess energy of the carriers acquired from $E$ is transferred to the phonons via electron–phonon interactions. As in the treatment used in the study of hot electrons in semiconductors discussed in the Introduction section, we assume that the distribution of electrons under $E$ is represented by a high-temperature Fermi distribution. Accordingly, we can model the generation of heat as the energy exchange between a "hot electron" system having electron temperature $T_e$ and a "cold phonon" system with lattice temperature $T_l$.

In this study, we derive the energy relaxation rate from the evolution of the energy of the phonon system by considering phonon excitation as the initial process of Joule heating. For simplicity, we approximate that the excited phonons decay instantaneously via phonon-phonon interactions into thermal phonons which immediately radiate/transport the thermal energies to the environment/heat bath. Note that the effect of electron–electron scattering is regarded as the source of electron equilibration and not explicitly treated in the evaluation of energy relaxation because of its much shorter time scale than phonon excitation [30]. The Boltzmann equation for the distribution of phonons with wavevector **q** can be expressed as follows:

$$\frac{\partial n_{\mathbf{q}}}{\partial t} = \left.\frac{\partial n_{\mathbf{q}}}{\partial t}\right|_{diff} + \left.\frac{\partial n_{\mathbf{q}}}{\partial t}\right|_{coll}, \quad (1)$$

where $n_{\mathbf{q}}$ represents the distribution function of the phonons, $\left.\frac{\partial n_{\mathbf{q}}}{\partial t}\right|_{diff}$ is the diffusion term induced by the temperature gradient, and $\left.\frac{\partial n_{\mathbf{q}}}{\partial t}\right|_{coll}$ is the collision term induced by the electron–phonon interaction. Here, we assume that the temperature gradient is negligibly small, and the change in the distribution function of the phonon system is determined by the collision term. The contribution of the inelastic scattering process of the phonons with wavevector $\mathbf{q}$ and band index $\nu$ to the collision term can be divided into the emission process, $P_{\mathbf{q}\nu}^{+}$, and the absorption process, $P_{\mathbf{q}\nu}^{-}$.

$$P_{\mathbf{q}\nu}^{+} = 2\frac{2\pi}{\hbar}\sum_{\mathbf{k}}|M^{nm\nu}(\mathbf{k},\mathbf{q})|^2(n_{\mathbf{q}\nu}+1)\delta(E_{n\mathbf{k}}-E_{m\mathbf{k}+\mathbf{q}}+\hbar\omega_{\mathbf{q}\nu})(1-f_{n\mathbf{k}})f_{m\mathbf{k}+\mathbf{q}}, \quad (2)$$

$$P_{\mathbf{q}\nu}^{-} = 2\frac{2\pi}{\hbar}\sum_{\mathbf{k}}|M^{nm\nu}(\mathbf{k},\mathbf{q})|^2 n_{\mathbf{q}\nu}\delta(E_{m\mathbf{k}+\mathbf{q}}-E_{n\mathbf{k}}-\hbar\omega_{\mathbf{q}\nu})(1-f_{m\mathbf{k}+\mathbf{q}})f_{n\mathbf{k}}, \quad (3)$$

where $(n,m)$ and $\mathbf{k}$ are the band indices and wavevector of the electrons, respectively. $M^{nm\nu}(\mathbf{k},\mathbf{q})$ is the electron–phonon coupling matrix element associated with the scattering process. The factor of two corresponds to the spin degrees of freedom in the electron system. The evolution of the phonon distribution is expressed as

$$\left.\frac{\partial n_{\mathbf{q}}}{\partial t}\right|_{coll} = \sum_{\nu} P_{\mathbf{q}\nu}^{+} - P_{\mathbf{q}\nu}^{-}$$
$$= 2\frac{2\pi}{\hbar}\sum_{\mathbf{k},\nu}|M^{nm\nu}(\mathbf{k},\mathbf{q})|^2\{n_{\mathbf{q}\nu}(f_{m\mathbf{k}+\mathbf{q}}-f_{n\mathbf{k}})+f_{m\mathbf{k}+\mathbf{q}}(1-f_{n\mathbf{k}})\}\delta(E_{n\mathbf{k}}-E_{m\mathbf{k}+\mathbf{q}}+\hbar\omega_{\mathbf{q}\nu}). \quad (4)$$

Thus, the rate of change in the energy in the phonon system by the collision term can be written as

$$\left.\frac{\partial E_{ph}}{\partial t}\right|_{coll} = 2\frac{2\pi}{\hbar}\sum_{\mathbf{k},\mathbf{q},\nu}\hbar\omega_{\mathbf{q}\nu}|M^{nm\nu}(\mathbf{k},\mathbf{q})|^2\{n_{\mathbf{q}\nu}(f_{m\mathbf{k}+\mathbf{q}}-f_{n\mathbf{k}})+f_{m\mathbf{k}+\mathbf{q}}(1-f_{n\mathbf{k}})\}\delta(E_{n\mathbf{k}}$$
$$-E_{m\mathbf{k}+\mathbf{q}}+\hbar\omega_{\mathbf{q}\nu}). \quad (5)$$

Hereafter, Eq. (5) is referred as the energy relaxation rate because it corresponds to the rate of the increase (decrease) in the energy in the phonon (electron) system.

B.  Relationship between electron temperature and applied electric field

Formulating the relation between a given $T_e$ and $E$ is necessary to express our calculation results as a function of $E$ and to compare the results with those of previous reports. This can be achieved by considering the balance between the energy relaxation rate and the macroscopic definition of Joule heating derived from the mobility calculation under the two-temperature concept. In this study, we consider that the electron–electron and electron–impurity scattering are the sources of electron equilibration, and their effects are described by a Fermi distribution with $T_e$. The energy transfer from electrons to phonons, that is, Joule heating is governed by inelastic electron–phonon scattering. Based on this approximation, we evaluated Joule heating using the phonon-limited mobility which is a function of both $T_e$ and $T_l$. Within the linearized Boltzmann equation (LBE) and relaxation time approximation (RTA), the two-temperature version of the electron mobility tensor $\mu_{\alpha\beta}(T_e, T_l)$ can be written as

$$\mu_{\alpha\beta}(T_e, T_l) = -\frac{e}{N_c\Omega}\sum_n \int \frac{d\mathbf{k}}{\Omega_{BZ}} \frac{\partial f_{n\mathbf{k}}(T_e)}{\partial E_{n\mathbf{k}}} v_{n\mathbf{k},\alpha} v_{n\mathbf{k},\beta} \tau_{n\mathbf{k}}(T_e, T_l), \quad (6)$$

where $N_c$ is the carrier concentration, $v_{n\mathbf{k},\alpha}$ is the band velocity of the $n$-th band at $\mathbf{k}$ along the $\alpha$ direction, and $\tau_{n\mathbf{k}}(T_e, T_l)$ is the electron–phonon relaxation time, which is defined as

$$\frac{1}{\tau_{n\mathbf{k}}(T_e, T_l)} = \frac{2\pi}{\hbar}\sum_{m\nu}\int \frac{d\mathbf{q}}{\Omega_{BZ}}|M^{nm\nu}(\mathbf{k},\mathbf{q})|^2 \left[\{1 - f_{m\mathbf{k}+\mathbf{q}}(T_e) + n_{\mathbf{q}\nu}(T_l)\}\delta(E_{n\mathbf{k}} - E_{m\mathbf{k}+\mathbf{q}} - \hbar\omega_{\mathbf{q}\nu}) \right.$$
$$\left. + \{f_{m\mathbf{k}+\mathbf{q}}(T_e) + n_{\mathbf{q}\nu}(T_l)\}\delta(E_{n\mathbf{k}} - E_{m\mathbf{k}+\mathbf{q}} + \hbar\omega_{\mathbf{q}\nu})\right]. (7)$$

Considering that $\left.\frac{\partial E_{ph}}{\partial t}\right|_{coll}$ in Eq. (5) represents the energy gain from $E$, this should be equivalent to Joule heating. Using the macroscopic definition of Joule heating in an isotropic crystal structure, the following equation holds for a given set of $T_e$ and $T_l$:

$$e\mu_{avg}(T_e, T_l)E^2 = \frac{1}{N_c}\left.\frac{\partial E_{ph}}{\partial t}\right|_{coll}. \quad (8)$$

Here, $\mu_{avg}$ is the average of the diagonal part of $\mu_{\alpha\beta}$ in Eq. (6). By calculating $\mu_{\alpha\beta}(T_e, T_l)$ and $\left.\frac{\partial E_{ph}}{\partial t}\right|_{coll}$ at the same $T_l$, we can evaluate the value of the electric field equivalent to $T_e$.

C. DFT calculations

We calculated the electron–phonon interaction and electron and phonon eigenstates in Eq. (5) using Quantum Espresso [31–33] and the Electron–Phonon Wannier (EPW) package [24,34,35]. We developed an extension of EPW to evaluate the energy relaxation rate and mobility under the two-temperature model. The electronic structure and phonon properties of a bulk Si crystal were calculated using a norm-conserving pseudopotential [36,37]. The exchange and correlation functional with the generalized gradient approximation of Perdew, Burke, and Ernzerhof [38] was used. The kinetic cutoff energy was set to 80 Ry. The lattice constant of the Si primitive cell was set to 5.48 Å. The Brillouin zone (BZ) was sampled on a $12 \times 12 \times 12$ ($6 \times 6 \times 6$) uniform mesh for the electronic state (phonon). The obtained electron and phonon eigenvalues and electron–phonon matrix elements were interpolated using maximally localized Wannier functions. In the interpolation, random grids of 200,000 **k**- and **q**- points were sampled to evaluate Eq. (5). With this dense sampling, the calculated electron (hole) mobility at 350 K was 912.9 (396.4) cm$^2$/Vs, which reproduces previously reported values well [23,24]. The concentrations of both the electron and hole carriers were set to $1.0 \times 10^{15}$

cm$^{-3}$. The bandgap is corrected by the rigid shift of the conduction band energy by 0.6 eV.

The validity of the calculation results of the energy relaxation rate was confirmed by decomposing the energy relaxation rate into phonon emission ($P_{\mathbf{q}\nu}^+$) and phonon absorption ($P_{\mathbf{q}\nu}^-$) processes (see Supplementary Materials Sec.1). The total energy relaxation rate approaches zero as $T_e = T_l$ by the cancellation of phonon emission and absorption processes, except for the small residual caused by discrete sampling of the reciprocal space. To suppress the effect of this residual, we used the results obtained at $T_e$ sufficiently higher than $T_l$; $T_e$= 350-950 (35 -105) K for $T_l$=300 (1) K.

We ignored the spin-orbit interaction (SOI) in the following calculations. The deformation of the band structure by SOI can affect the relaxation process of the carriers especially at a low $T_l$. This point is discussed in the Supplementary Materials Sec. 5.

III.     Results and Discussion

A. Conversion from electron temperature to applied field

For bulk Si, we calculated $\mu_{\alpha\beta}(T_e, T_l)$ and $\left.\frac{\partial E_{ph}}{\partial t}\right|_{coll}$ at $T_l$ = 1 and 300 K, respectively. We then evaluated the value of $E$ equivalent to $T_e$ based on Eq. (8). Fig. 1 shows the results of this calculation. In what follows, we converted $T_e$ to $E$ based on the results shown in Fig. 1.

The compatibility between $T_e$ and $E$ is a central concept in our framework. To verify this point, we also evaluated the dependence of the mobility and drift velocity $v_d$ on $E$ at $T_l$ =300 K, and compared the results with those of the previous studies. Fig. 2 (a) shows the calculation results of $\mu_{avg}(E, T_l = 300 \text{ K})$. According to a previous report, in Si, the carrier mobility as a function of $E$ is phenomenologically expressed as [39]

$$\mu_{avg}(E) = \frac{\mu_0}{\left[1 + \left(\frac{E}{E_c}\right)^\beta\right]^{\frac{1}{\beta}}}. \quad (9)$$

Here $\mu_0$ (cm$^2$/Vs), $E_c$ (kV/cm), and $\beta$ are the fitting parameters. As shown by the continuous lines in Fig 2 (a), the best fit of $\mu_{avg}(E, T_l = 300K)$ for electrons (holes) was obtained using Eq. (9) by $\mu_0 = 1262.3$ (536.0), $E_c = 9.56$ (23.84), and $\beta = 1.55$ (1.75).

For the estimation of $v_d$, we used the following equation:

$$e\mu_{avg}(E, T_l = 300K)E^2 = eE\, v_d\,. \quad (10)$$

In previous studies of high field effects in Si [39, 40], it was reported that the electron (hole) drift velocity in bulk Si is saturated at an electric field of 30–40 kV/cm (~100 kV/cm) at 300 K. The saturated value of the electron (hole) drift velocity is $1.0$–$1.1 \times 10^7$ cm/s ($0.95$–$1.0 \times 10^7$ cm/s). As shown in Fig. 2 (b), the dependence of the drift velocities on $E$ reproduced the general tendency of the experimental observations. The calculation result of the nearly saturated drift velocity in the electron carrier ($1.1 \times 10^7$ cm/s at 34 kV/cm) shows reasonable agreement with previous results, but that in the hole carrier ($1.2 \times 10^7$ cm/s at 72 kV/cm) shows an overestimation. This may be attributed to the stronger influences of the following factors in a higher electric field: limitations of the LBE and RTA treatments, and approximation of electron-electron and electron-impurity scattering.

B. Microscopic detail of energy relaxation rate

First, we investigated the general trends of the energy relaxation rate for low $T_l = 1$ K and high $T_l = 300$ K cases. Fig. 3 shows the calculation results of the energy relaxation rate as a function of $T_e$ and $E$. The calculated energy relaxation rate in the high-temperature case is in agreement with a previously reported result. Based on a Monte Carlo simulation using empirical pseudopotentials, the energy loss rate per electron carrier was estimated as ~0.4 eV/ps at $E = 30$ kV/cm [5]. Converting our energy relaxation rate per unit cell to that per electron carrier yields a value of 0.352 eV/ps at $E = 32$ kV/cm.

The notable difference between the high and low lattice temperature cases appears in the dependence of the energy relaxation rate on $E$. At a high $T_l$, the energy relaxation rate $\propto E$, whereas it is $\propto E^2$ at a low $T_l$. The origin of this behavior is rationalized by the combination of the contributions of the acoustic and optical phonons. At a low $T_l$, the thermal excitation of the optical phonons is suppressed. When $T_l$ is less than the optical phonon energy, the acoustic phonons play a dominant role in the energy transfer from the carriers to the lattice system. In bulk Si, optical phonon branches exist at approximately 50–60 meV. Considering the Bose distribution, thermal excitation of these optical phonons occurs when $T_l > 60$ K. At $T_l = 1$ K, the thermal excitation is suppressed. Instead, hot electrons with $T_e > 60$ K ($E$ ~50 V/cm for the electron carriers and ~ 100 V/cm for the hole carriers) excite the optical phonons. The presence of the threshold voltage for optical phonon excitation results in a nonlinear dependence on $E$. In contrast, at a high $T_l$, thermal excitation of optical phonons can occur, which makes the energy relaxation rate less sensitive to $E$.

The above changeover behavior of the relaxation process was theoretically discussed and experimentally observed in two–dimensional electron gas (2DEG) systems emerging in a Si inversion layer, GaAs/AlGaAs interfaces, and GaN/AlGaN heterojunctions [2,41–45]. However, the band structure of bulk Si is different from that of the 2DEG system, and Si is a nonpolar crystal without polar electron–phonon coupling, which is dominant in the latter heterojunction. To confirm that the changeover originated from the optical phonons, we decomposed the energy relaxation process. As shown in Fig. 4 (a), the energy range of the phonons is divided into three regions. We set the thresholds to 27 meV and 50 meV. The phonons with <27 meV are the longitudinal acoustic (LA) phonons around the zone center and the transverse acoustic (TA) mode. The second energy range between 27 and 50 meV includes the contribution from the LA mode with a finite wavevector and the longitudinal optical (LO) mode at the zone boundary. The highest energy range of >50 meV corresponds to the contribution from the transverse optical (TO) modes and the LO mode around the zone center.

Figs. 4 (b)–(d) show the decomposition of the energy relaxation rates into the contributions from the phonons with <27 meV, <50 meV, and >50 meV energy at low and high $T_l$ as functions of $E$. We found that the LA phonons around the $\Gamma$ point and the TA modes make a minor contribution to the energy relaxation processes under all conditions: low and high $T_l$ and $E$. At low $T_l$, the TO and the zone center LO modes with >50 meV energy contribute less at low $E$, whereas they become dominant as $E$ increases. This behavior corresponds to the changeover behavior discussed above. The contributions from the phonons with medium energy, the LA and TO modes at the zone boundary, differ for the electron and hole carriers. For the electron carriers, there is a substantial contribution from medium-energy phonons at both low and high $T_l$. In particular, the relaxation via medium-energy phonons is dominant at low $E$ at low $T_l$. In contrast, for the hole carriers, the contribution from medium-energy phonons is minor compared to that from the TO and zone center LO modes, except at a very low $E$ at a low $T_l$.

The difference in the energy relaxation processes of the electron and hole carriers becomes evident in Fig.5, where we plot the relaxation rate at each **k**-point defined by

$$\left.\frac{\partial E_{ph}(E_{n\mathbf{k}})}{\partial t}\right|_{coll} = 2\frac{2\pi}{\hbar}\sum_{\mathbf{q},\nu}\hbar\omega_{\mathbf{q}\nu}|M^{nm\nu}(\mathbf{k},\mathbf{q})|^2\{n_{\mathbf{q}\nu}(f_{m\mathbf{k}+\mathbf{q}}-f_{n\mathbf{k}})+f_{m\mathbf{k}+\mathbf{q}}(1-f_{n\mathbf{k}})\}$$
$$\delta(E_{n\mathbf{k}}-E_{m\mathbf{k}+\mathbf{q}}+\hbar\omega_{\mathbf{q}\nu}). \quad (11)$$

$E_{n\mathbf{k}}$ is the energy of the final state for the phonon emission process and that of the initial state for the phonon absorption case. Here, we focus on the $T_e > T_l$ case, in which the phonon emission process is dominant. Because $E_{n\mathbf{k}}, E_{n\mathbf{k}+\mathbf{q}} \gg \hbar\omega_{\mathbf{q}\nu}$, we considered $E_{n\mathbf{k}}$ as the carrier energy. For the electron carriers, the energy transfer from the electrons to the phonons is maximized at the CBM, and it monotonically decreases as the carrier energy increases from the CBM. In contrast, for the hole carriers, the energy relaxation rate is low at the VBM, whereas, it is enhanced at by approximately 100 meV below the VBM. As discussed below, the above distinct difference in the energy relaxation rate distribution of the electron and hole carrier energies can be explained by the differences in the dominant electron scattering by the phonons around the CBM and the VBM.

In bulk Si, the CBM is located at a $\Delta_1$ valley close to the X point in the first BZ. Because of the symmetry in the diamond crystal structure, there are six positions equivalent to the $\Delta_1$ valley. Thus, intervalley scattering can occur via electron–phonon interactions. Owing to the symmetry constraint, the intervalley scattering is mediated by the LO phonons (g-process) and LA or TO phonons (f-process) [46,47]. Intravalley scattering caused by optical phonons is forbidden. In contrast, the VBM is positioned at the Γ point; thus, intervalley scattering process does not take place. Instead, intravalley and interband scattering can occur between the three bands constructing the VBM.

The projection of the electron and hole energy relaxation rates onto the band structure supports the above discussion (Fig. 6). In Fig. 6, $T_l$ is set to 300 K and $T_e$ is set to 600 K for both the electron and hole carrier calculations. For the electron carriers, the energy relaxation mostly occurs at the CBM located at the $\Delta_1$ valley, as expected from Fig. 5 and the above discussion. In contrast, for the hole carriers, the energy relaxation at the VBM is minor and maximized at a specific **k**-point region. In this study, we neglect the SOI; thus, the three bands, conventionally referred to as heavy-hole, light-hole, and split-off bands, are degenerate at the Γ point. The enhancement at the specific **k**-point region was the most clear in the highest heavy-hole band. Several possible explanations exist for this non-monotonic relaxation rate distribution of the hole carriers. One is the minor contribution of the lower-energy phonons in the hole carriers. As shown in Fig. 4, the energy relaxation of a hole carrier is governed by optical phonons in the range of >50 meV. This suppresses the energy relaxation within 50 meV below the VBM and shifts the overall energy relaxation rate toward the lower energy side. The other factor is the anisotropic band structure of the valence band. It is reported that the Si valence band dispersions are highly anisotropic compared to the conduction band, and are known as the "warping" band structure. This warping is stronger in the heavy-hole band than in the light-hole and split-off bands [47,48], as is visualized by the isoenergetic surface in Fig. S3. The projection of the energy relaxation rate at 100 meV below the VBM on the $k_z = 0$ plane in the reciprocal space shows that the apex areas of the warped heavy-hole band largely contribute to the energy relaxation of the hole (Fig.7). This indicates that the characteristic valence-band anisotropy imposes additional constraints on the electron and phonon wavevectors involved in the electron–phonon scattering process. This is also supported by the comparison of the energy relaxation rates of electrons and holes projected on the phonon band structure (Fig. S4). These additional energy and momentum constraints of the electron–phonon scattering process explain the differences in the energy relaxation rates of the electron and hole carriers.

IV.     Conclusion

We developed a theoretical framework for ab initio evaluation of the carrier energy relaxation in semiconductors under an applied electric field by combining the two-temperature model and the Boltzmann equation. We employed this method to investigate the initial process of Joule heating for n- and p-doped bulk Si. We found that the microscopic initial process of Joule heating differed in the electron and hole carriers. Phonons with 27–50 meV energy, LA phonons, and optical phonons at the zone boundary contribute to the energy relaxation process of the electron carriers, whereas they contribute negligibly to that of the hole carriers. Moreover, at the band edge, the energy relaxation rate is maximized in the electron carriers, whereas it is suppressed in the hole carriers. Our calculation results of the momentum-resolved energy relaxation rate indicate that the origin of the above-mentioned differences can be attributed to the presence/absence of the intervalley scattering process and the isotropic/anisotropic band structures in the electron and hole carriers. The important factors that govern the energy relaxation process—the intervalley scattering process in the electron carriers and the warped band structure in the hole carriers—can be controlled by strain [49,50]. A detailed ab initio investigation of the energy relaxation rate in a strained structure remains a topic for future research.

Acknowledgements

This work was supported by JST, PRESTO, Grant Number JPMJPR17I7, Japan. E. M. thanks N. Takagi for constructive discussions, and K. Hinode and S. Okugawa for supporting the analysis of the simulation results.

222104 (2019).

[6]   S. Das Sarma, J. K. Jain, and R. Jalabert, *Many-Body Theory of Energy Relaxation in an Excited-Electron Gas via Optical-Phonon Emission*, Phys. Rev. B **41**, 3561 (1990).

[7]   J. Sjakste, K. Tanimura, G. Barbarino, L. Perfetti, and N. Vast, *Hot Electron Relaxation Dynamics in Semiconductors: Assessing the Strength of the Electron–Phonon Coupling from the Theoretical and Experimental Viewpoints*, J. Phys. Condens. Matter **30**, 353001 (2018).

[8]   H. Tanimura, J. Kanasaki, K. Tanimura, J. Sjakste, and N. Vast, *Ultrafast Relaxation Dynamics of Highly Excited Hot Electrons in Silicon*, Phys. Rev. B **100**, 35201 (2019).

[9]   J. Sjakste, N. Vast, G. Barbarino, M. Calandra, F. Mauri, J. Kanasaki, H. Tanimura, and K. Tanimura, *Energy Relaxation Mechanism of Hot-Electron Ensembles in GaAs: Theoretical and Experimental Study of Its Temperature Dependence*, Phys. Rev. B **97**, 64302 (2018).

[10]  P. B. Allen, *Theory of Thermal Relaxation of Electrons in Metals*, Phys. Rev. Lett. **59**, 1460 (1987).

[11]  T. Low, V. Perebeinos, R. Kim, M. Freitag, and P. Avouris, *Cooling of Photoexcited Carriers in Graphene by Internal and Substrate Phonons*, Phys. Rev. B **86**, 45413 (2012).

[12]  P. Maldonado, K. Carva, M. Flammer, and P. M. Oppeneer, *Theory of Out-of-Equilibrium Ultrafast Relaxation Dynamics in Metals*, Phys. Rev. B **96**, 174439 (2017).

[13]  S. Sadasivam, M. K. Y. Chan, and P. Darancet, *Theory of Thermal Relaxation of Electrons in Semiconductors*, Phys. Rev. Lett. **119**, 136602 (2017).

[14]  L. Waldecker, R. Bertoni, R. Ernstorfer, and J. Vorberger, *Electron-Phonon Coupling and Energy Flow in a Simple Metal beyond the Two-Temperature Approximation*, Phys. Rev. X **6**, 21003 (2016).

[15]  Z. Lu, A. Vallabhaneni, B. Cao, and X. Ruan, *Phonon Branch-Resolved Electron-Phonon Coupling and the Multitemperature Model*, Phys. Rev. B **98**, 134309 (2018).
16

Figure captions

Fig. 1

Relationship between $T_e$ and $E$ calculated by using Eq. (8) at (a) $T_l$ = 300 K and (b) $T_l$ = 1 K. Because of the difference between the conduction and valence band dispersions, a slight difference exists between $E$ of the electron and hole carriers for the same values of $T_e$.

Fig. 2

(a) Mobilities and (b) drift velocities of electron and hole carriers calculated as a function of $E$ at $T_l$ =300K. The continuous lines in (a) are the fitting results obtained with Eq. (9).

Fig. 3

Energy relaxation rates of electron and hole carriers calculated as a function of $T_e$ at (a) $T_l$ = 300 K and (b) $T_l$ = 1 K. For completeness, the results are also plotted as a function of $E$.

Fig. 4

(a) Phonon band structure of bulk Si. The background colors indicate the three energy ranges used in (b) and (c). The energy relaxation rates of electron and hole carriers are decomposed into contributions from phonons in the three energy ranges: below 27 meV (red dotted), from 27 to 50 meV (blue solid), and above 50 meV (black double dotted) at (b) $T_l$ = 300 K and (c) $T_l$ = 1 K.

Fig. 5

Energy relaxation rates of (a) electron and (b) hole carriers calculated for each **k**-point and carrier energy with Eq. (11). Red (blue) dots correspond to the results at $T_e$ = 900 (600) K with $T_l$ = 300 K.

Black solid lines at ~ 0.6 eV in (a) and ~ –0.6 eV in (b) show the positions of CBM and VBM, respectively. Zero energy is located in the middle of VBM and CBM.

Fig. 6

Momentum dependence of the energy relaxation rates of (a) electron and (b) hole carriers. The energy relaxation rate calculated at each **k**-point is projected on the electron band structure around the CBM and VBM. The rates are represented by the color bar in each figure.

Fig. 7

Projection of energy relaxation rates of hole carriers at 100 meV below VBM on the $k_z = 0$ plane in the reciprocal space. The projection value is calculated by smearing the original data with a Gaussian function whose half width at half maximum is set to be $\sqrt{2\ln 2} \cdot 20$ meV. The white lines indicate the boundary of the first BZ. The rates are represented by the color bar in each figure.

Fig. 1

(a) Lattice temperature: 300 K

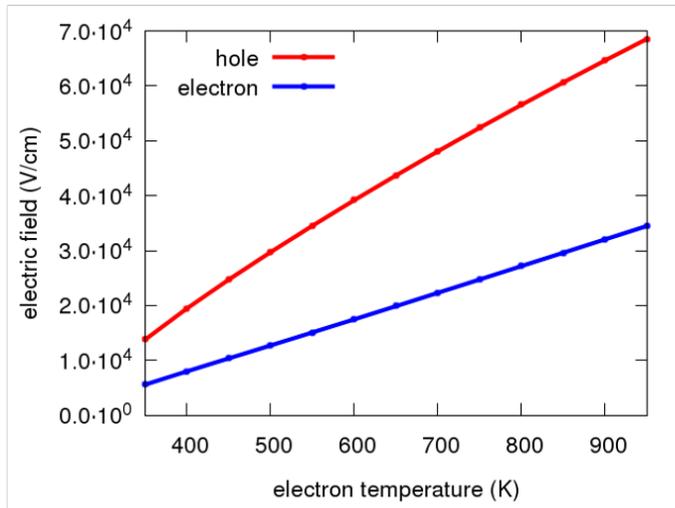

(b) Lattice temperature: 1 K

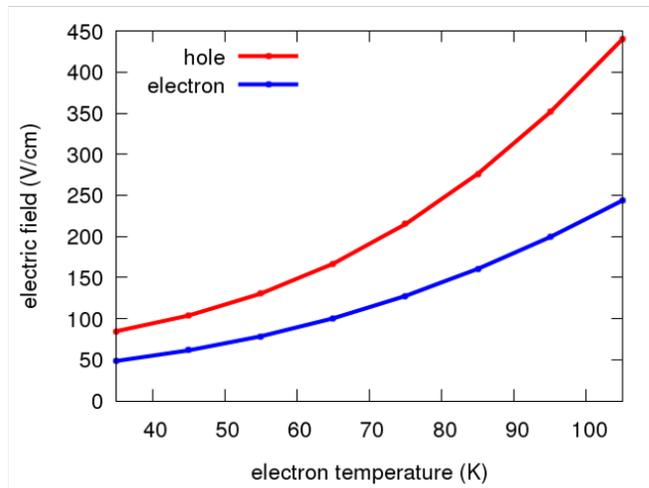

Fig 2

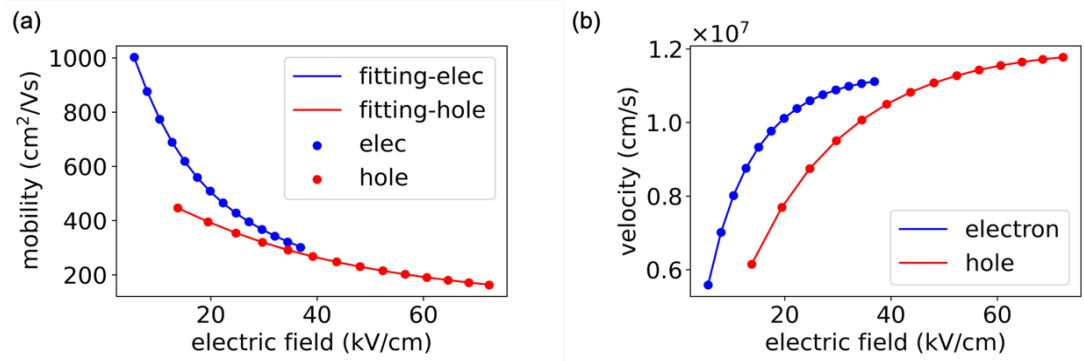

Fig. 3

## (a)    Lattice temperature: 300 K

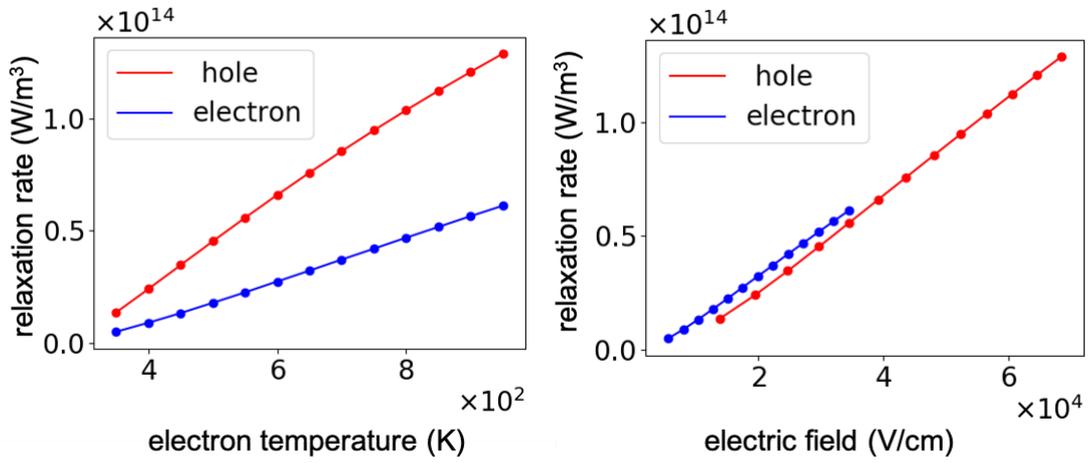

## (b)    Lattice temperature: 1 K

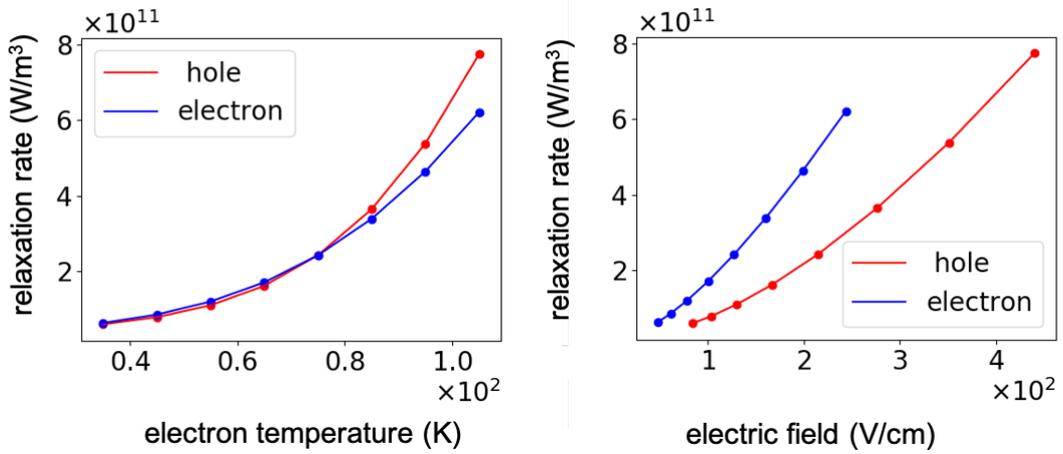

Fig. 4

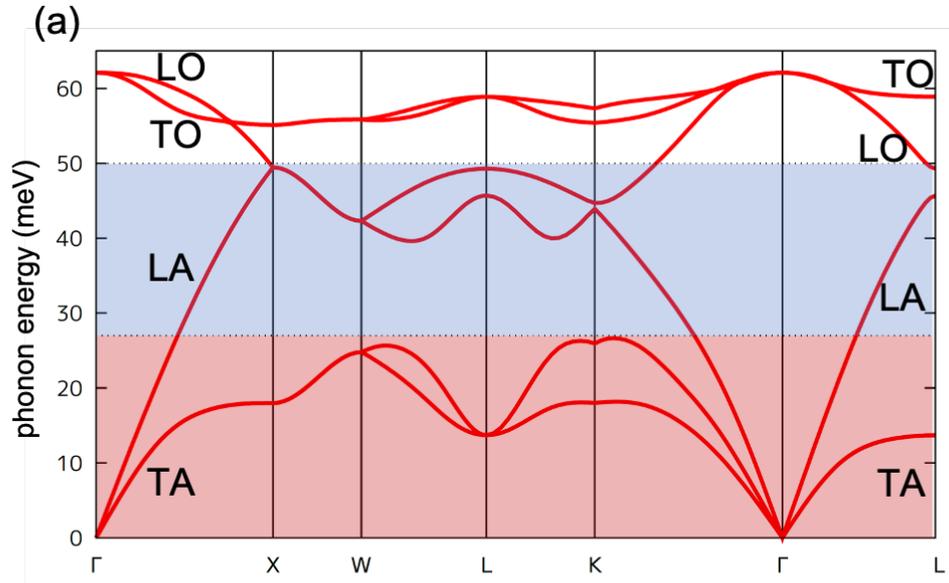

(a)

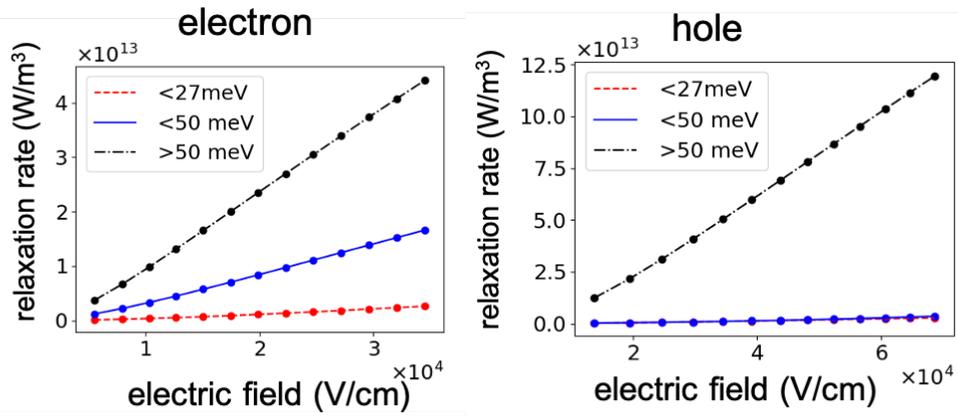

(b) $T_l$ = 300 K

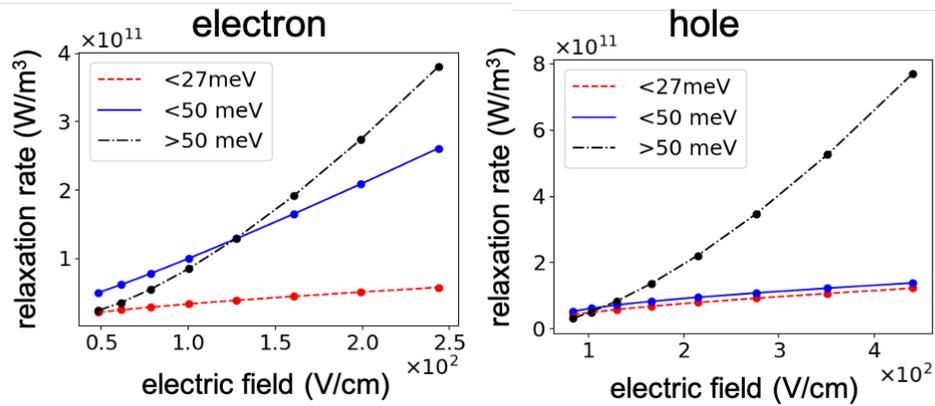

(c) $T_l$ = 1 K

Fig. 5

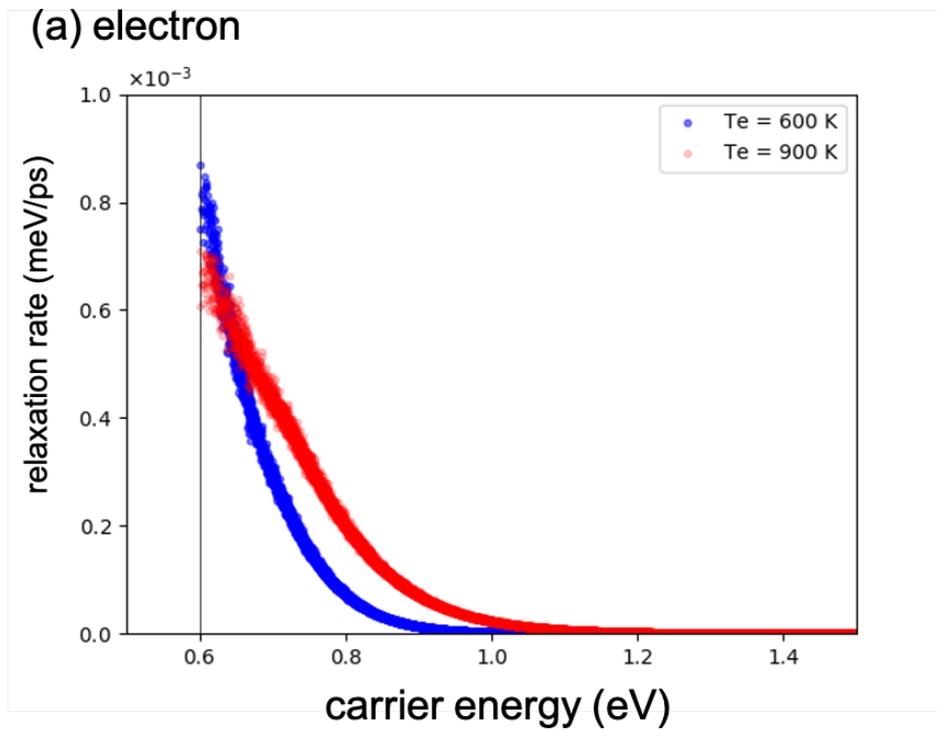

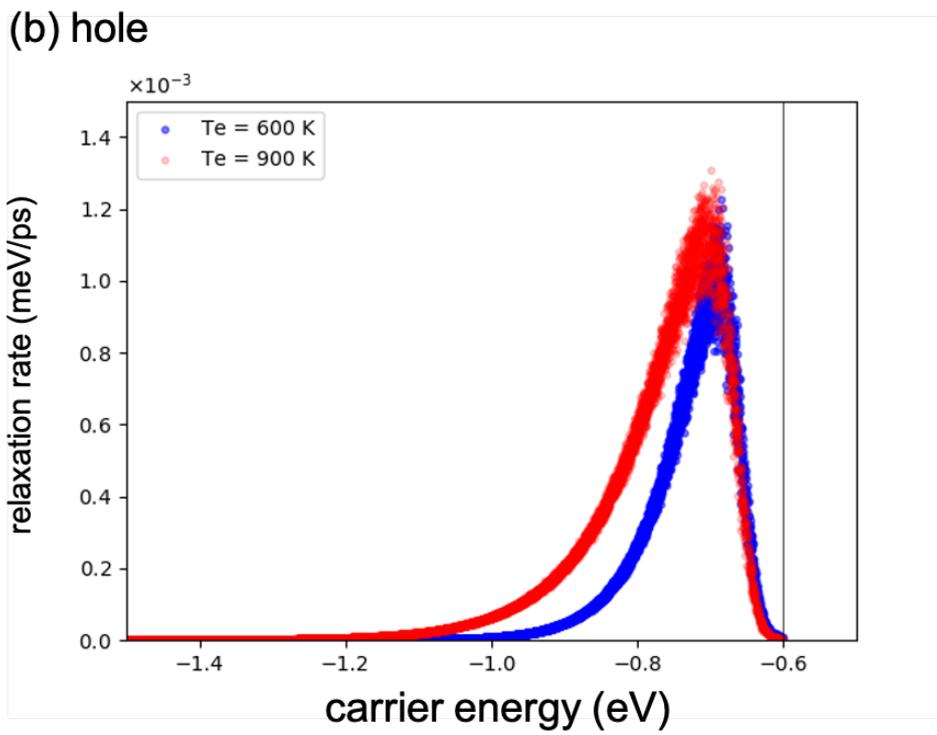

Fig. 6

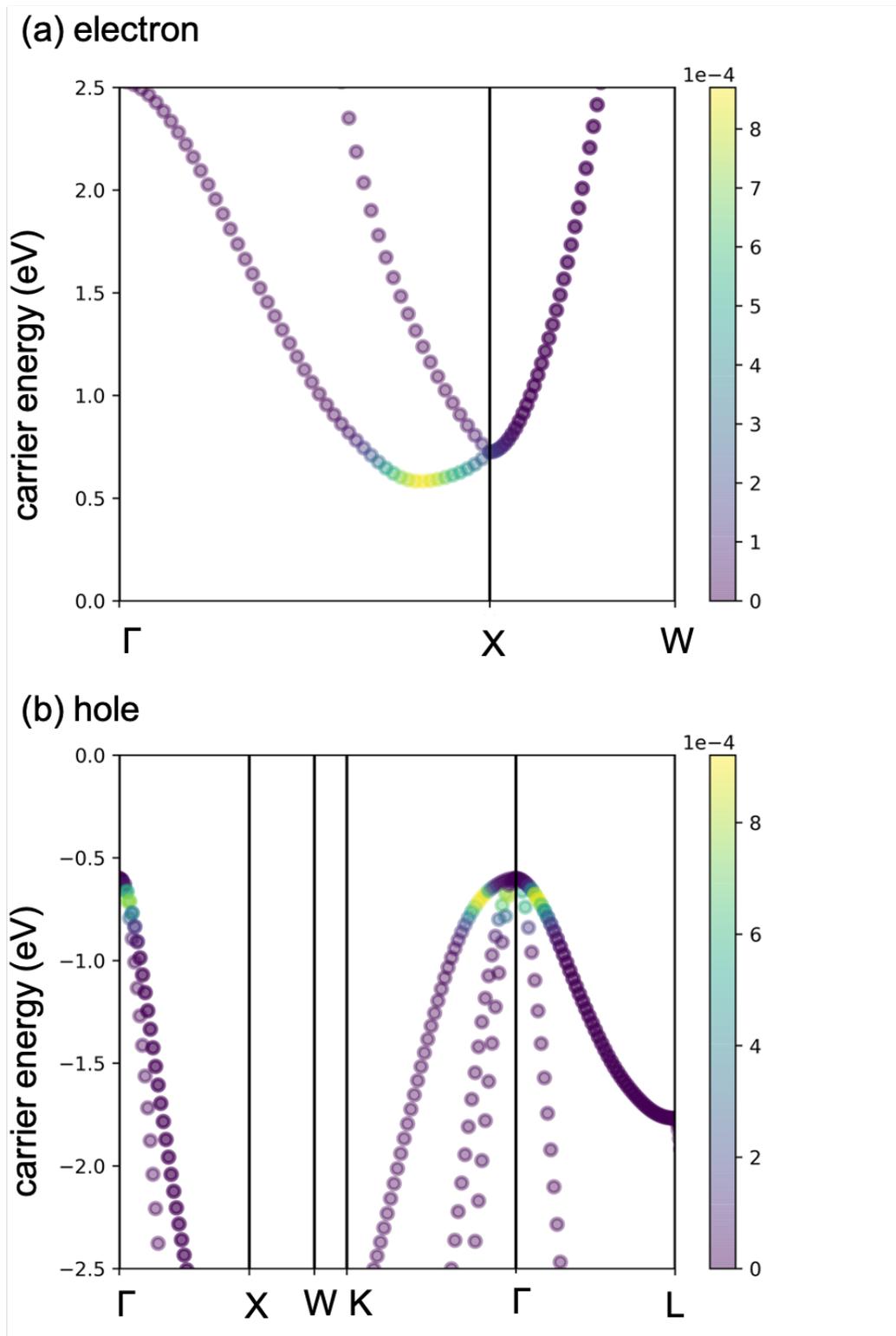

Fig.7

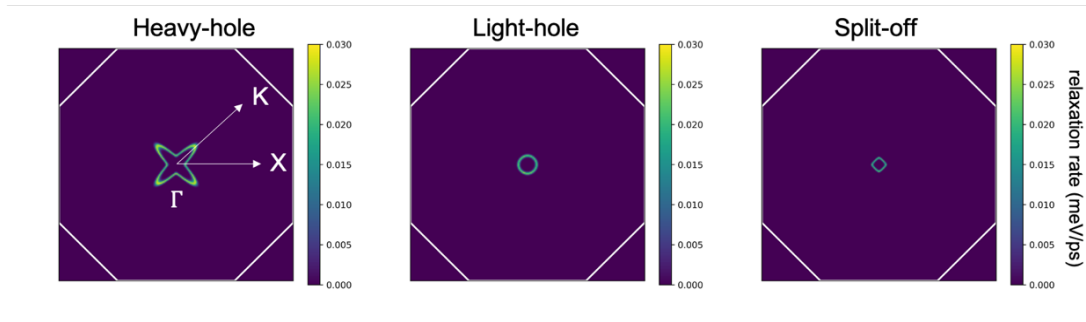